\documentstyle[epsf]{l-aa}
\def \aj#1#2   {AJ  {\bf #1}, #2}  
\def \apj#1#2  {ApJ {\bf #1}, #2}  
\def \apjs#1#2 {ApJ Suppl.Ser. {\bf #1}, #2}
\def \apss#1#2 {ApSS {\bf #1}, #2}
\def \aua#1#2  {A\&A {\bf #1}, #2}
\def \auas#1#2 {A\&AS {\bf #1}, #2} 
\def \araa#1#2 {ARAA {\bf #1}, #2} 
\def \azh#1#2  {Astr.Zh. {\bf #1}, #2}  
\def \iau#1#2  {IAU Symp. {\bf #1}, #2}
\def \mn#1#2   {MNRAS  {\bf #1}, #2} 
\def \pasj#1#2 {PASJ  {\bf #1}, #2}
\def \pasp#1#2 {PASP  {\bf #1}, #2}
\def \pnac#1#2 {Proc.\allowbreak Nat.\allowbreak Acad.\allowbreak Sci.
                 {\bf #1}, #2}
\def \rmaa#1#2 {Rev.\allowbreak Mex.\allowbreak A\&A {\bf #1}, #2}

\newcommand{\msun}{{\rm M}_\odot}
\newcommand{\Msun}{M$_\odot$}

\newif \ifkiel
\kieltrue 
\ifkiel
  \newcommand{\gdot}{\raisebox{0.25ex}{$\stackrel{.}{g}$}}
  \newcommand{\cldot}{\raisebox{0.25ex}{$\stackrel{.}{c}$}} 
  \newcommand{\sdot}{\raisebox{0.25ex}{$\stackrel{.}{s}$}}  
  \newcommand{\rdot}{\raisebox{0.25ex}{$\stackrel{.}{r}$}}  
  \newcommand{\egdot}{\raisebox{0.25ex}{$\stackrel{.}{e}_g$}}
  \newcommand{\tgdot}{\raisebox{0.25ex}{$\stackrel{.}{T}_g$}}     
  \newcommand{\ecdot}{\raisebox{0.25ex}{$\stackrel{.}{e}_c$}} 
  \newcommand{\tcdot}{\raisebox{0.25ex}{$\stackrel{.}{T}_c$}} 
  \newcommand{\mcdot}{\raisebox{0.25ex}{$\stackrel{.}{M}_c$}} 
  \newcommand{\zcdot}{\raisebox{0.25ex}{$\stackrel{.}{z}_c$}} 
  \newcommand{\zgdot}{\raisebox{0.25ex}{$\stackrel{.}{z}_g$}} 
  \newcommand{\zsdot}{\raisebox{0.25ex}{$\stackrel{.}{z}_s$}} 
  \newcommand{\zrdot}{\raisebox{0.25ex}{$\stackrel{.}{z}_r$}} 
\else
  \def \gdot  {\dot{g}} 
  \def \cldot {\dot{c}} 
  \def \sdot  {\dot{s}} 
  \def \rdot  {\dot{r}} 
  \def \egdot {\dot{e}_g} 
  \def \tgdot {\dot{T}_g} 
  \def \ecdot {\dot{e}_c} 
  \def \tcdot {\dot{T}_c} 
  \def \mcdot {\dot{M}_c} 
  \def \zcdot {\dot{z}_c} 
  \def \zgdot {\dot{z}_g} 
  \def \zsdot {\dot{z}_s} 
  \def \zrdot {\dot{z}_r} 
\fi
\begin{document}
\thesaurus{3(09.01.1; 09.07.1; 10.05.1; 11.01.1; 11.05.2)}
\title{Condensation and evaporation of interstellar clouds in
       chemodynamical models of galaxies}

\author{J. K\"oppen \inst{1,2,3} 
     \and 
       Ch. Theis \inst{1}
     \and 
       G. Hensler \inst{1}}
\institute{Institut f\"ur Astronomie und Astrophysik der Universit\"at,
           D--24098 Kiel, Germany
       \and 
           URA 1280, 
           Observatoire de Strasbourg, 11 rue de l'Universit\'e, 
           F--67 000 Strasbourg, France
       \and
           International Space University,
           Parc d'Innovation, F-67 400 Illkirch, France}

\offprints{J. K\"oppen}
\maketitle

\begin{abstract}
    The network of interactions between hot gas, cool clouds, massive
    stars, and stellar remnants used in the chemodynamical modeling of the
    interstellar medium is investigated for the types of its solutions.
    In a physically consistent formulation for the energy transfer during
    condensation, oscillations due to a cyclic switching between condensation
    and evaporation never occur. A closed-box system evolves in a hierarchy
    of equilibria: thermal balance in the cloud gas, star-formation
    self-regulated due to heating of the clouds by massive stars, and
    the balance of condensation and evaporation. Except for an initial
    transitory phase, the evolution of the metallicity follows that of
    the Simple Model quite closely. For galaxies with a high initial
    density, or if the condensation rate is low, the metals produced
    by the stars may remain stored in the hot gas phase even in evolved
    systems with low gas fractions. 
\end{abstract}


              \section{Introduction}

    To understand the evolution of galaxies one may attempt to
    match the observational data by models which describe the global 
    processes -- star-formation rate, gas infall, gas loss -- with 
    suitable formulations. By adjusting free parameters, quite good
    fits can be achieved. However, the number of these free parameters
    often is uncomfortably large. Moreover, this approach may not lead
    to a unique identification of the dominant physical process, as
    the persisting G dwarf-`problem' (Pagel \& Patchett 1975) and
    the formation of radial abundance  gradients (G\"otz \& K\"oppen 1992)
    illustrate.

    Our chemodynamical approach (Hensler 1987, Theis et al. 1992,
    Samland et al. 1997) tries to describe as precisely as possible the
    known physical processes present in the interstellar medium (ISM) and 
    its interaction with the stars. These local processes are coupled
    with the global dynamics of the gaseous and stellar components,
    constituting a physical description of the evolution of a galaxy.
    Since it is unrealistic to include all processes in their full 
    complexity, one has to define a sufficiently simple but accurate 
    network of interactions in the ISM. Our prescription, based on the
    three-component ISM model of McKee \& Ostriker (1977) and on the
    formulations of Habe et al. (1981) and Ikeuchi et al. (1984),
    has successfully been coupled with the global dynamics for models
    of elliptical (Theis et al. 1992) and disk galaxies (Samland et
    al. 1997).

    Another important aspect is the degree of non-linearity of the 
    network which determines the behaviour of the model. Since the 
    dependences of the rate coefficients are not well known, some 
    caution is necessary to avoid the appearance of complex behaviour 
    solely due to the mathematical formulation.
    We cannot yet fully settle these questions, but what is needed is 
    a more complete understanding of the behaviour of this type of
    model and an identification of the crucial processes. 

    In their chemodynamical models  Theis et al. (1992) find that 
    most often the star-formation rate varies slowly with time, but 
    under certain conditions it undergoes strong nonlinear oscillations, 
    involving the condensation and evaporation of the cool clouds 
    embedded in the hot intercloud gas. The phases of slow variation are
    due to an equilibrium caused by the self-regulation of the
    star-formation rate (SFR) whose efficiency is reduced as the massive
    stars heat the gas by their ionizing continuum radiation. In a partial
    network with a single gas phase K\"oppen et al. (1995) show that this
    equilibrium results in a quadratic dependence of the SFR on gas
    density -- independent of what was assumed for the stellar birth
    function. Under realistic conditions this is quickly reached, and it
    is unconditionally stable, quite insensitive to the rate coefficients
    used. 

    The present study extends the network of K\"oppen et al. (1995)
    to two gas components, clouds and intercloud gas, described in Sect. 2. 
    We investigate its behaviour by numerical solution which allows
    the extraction of analytical conditions and relations.
    This permits the identification of the origin of the oscillations 
    of the SFR (Sect. 3), and the formulation of a physically consistent 
    description (Sect. 4) which leads to the identification of a second
    equilibrium, namely that between condensation and evaporation of 
    the clouds. In Sect. 5 we extend the prescription to the more 
    realistic one by Samland et al. (1997), having condensation and 
    evaporation occurring simultaneously in a cloud population.

          \section{The model for the ISM}
     
         \subsection{The description used in CDE models}

   We shall consider a somewhat simplified version of the present CDE
   models which captures their characteristic behaviour. As in the full
   models, there are four components: the hot intercloud gas (named
   hereafter `gas', with a mass density $g$), the gas in the form of
   clouds (`clouds' $c$), as well as massive stars ($s$), and low mass
   stars and remnants ($r$). Between the components the following
   interactions are taken into account: star-formation, gas return from
   dying stars, evaporation of clouds, condensation of gas onto clouds,
   (radiative or mechanical) heating of the gas by massive stars, radiative
   cooling of the gas. The full network also includes other processes,
   such as the formation of clouds by compression in supernova shells,
   dissipation by cloud-cloud collisions. These will not be included in
   our investigation, because comparison with the results of the complete
   network showed that they do not essentially determine the type of the
   system's behaviour. Then the time evolution of the mass
   densities of the components is described by the following equations:
   \begin{eqnarray}
       \gdot  & = & s \eta /\tau + E c - K g  \label{e:gas}   \\  
       \cldot & = & -\Psi        - E c + K g  \label{e:cloud} \\
       \sdot  & = & \xi \Psi - s/\tau         \label{e:stars} \\ 
       \rdot  & = & (1-\xi)\Psi + (1-\eta)s/\tau 
   \end{eqnarray}
   Throughout the paper, we shall use the units parsec, $10^6$ years,
   and solar masses.
   
   Star-formation is described by the stellar birth function 
   used in the form of K\"oppen et al. (1995)
   \begin{equation} 
      \label{e:sfr}
      \Psi(c,T_c) = C_n \, c^n \, \exp(-T_c/1000\,{\rm K})
   \end{equation} 
   Normally we use a quadratic dependence on density ($n=2$ and
   $C_2 = 0.55$). The exponential factor involving the temperature 
   $T_c$ of the cloud gas describes what fraction of a cloud is in 
   the form of star forming molecular clumps. 

   The mass returned to the interstellar gas by dying massive 
   stars (with a mean life-time $\tau = 10$ Myr) is taken to be the 
   fraction $\eta = 0.9$ of the stellar mass.
   Of all stars born, the fraction $\xi = 0.1$ is in the
   form of massive stars.

   The remaining terms pertain to evaporation of clouds, whose 
   rate coefficient $E$ can be a function of densities and 
   temperatures, and condensation of gas onto clouds (coefficient
   $K$).

   In the formulations of Hensler \& Burkert (1991) and Theis 
   et al. (1992) the cloudy medium is composed of clouds which 
   have identical properties (radius $R_c$, mass $M_c$, 
   density $\rho_c$) and which are embedded in the (hot) intercloud 
   gas. One assumes pressure equilibrium 
   ($T_c / \mu_c\epsilon_c =  g T_g / \mu_g\epsilon_g$), 
   which gives for the volume filling factor of the cloudy medium:
   \begin{equation} 
      \label{e:filling}
      \epsilon_c = {c T_c /\mu_c \over g T_g/\mu_g + c T_c/\mu_c} 
   \end{equation}
   with the mean molecular masses $\mu_c$, $\mu_g$ in the two gas 
   phases. For simplicity, we set $\mu_c = \mu_g = \mu_p$ equal to 
   one proton mass, in all what follows. The true mean density 
   $\rho_c$ within a single cloud is given by
   \begin{equation}
      \label{e:clouddens}
      \rho_c = c / \epsilon_c
   \end{equation}
   Assuming that the clouds are just unstable to the Jeans criterion,
   the mass is determined by
   \begin{eqnarray}
      \label{e:cloudmass}
      M_c = {4 \pi \rho_c \over 3} R_c^3
          = \sqrt{{3 \over 4 \pi \rho_c}} 
            \left( {k_{\rm B} T_c \over 0.2 G \mu_c}\right)^{1.5}  
                         \nonumber
   \end{eqnarray}
   with Boltzmann's constant $k_{\rm B}$ and gravitational constant 
   $G$. Hence the radius of the cloud is
   \begin{eqnarray}
      \label{e:cloudradius}
      R_c & = & \sqrt{ {3 \over 4 \pi} 
            {k_{\rm B} T_c \over 0.2 G \mu_c \rho_c} }       \\
      R_c [{\rm pc}] & = & 298.0 \, \cdot (T_c/10^4\,{\rm K})^{1/2}
              \cdot (\rho_c  [\msun /{\rm pc}^3] )^{-1/2} 
                        \nonumber 
   \end{eqnarray}
   Depending on its properties, such a cloud evaporates into the
   surrounding medium due to thermal conduction from the ambient
   hot gas or the gas condenses onto it. Cowie et al. (1981) 
   give a criterion for this behaviour: If the quantity defined as 
   \begin{eqnarray}
      \label{e:sigma}
      \sigma   & = & \left({T_g \over 1.54\, 10^7 {\rm K}}\right)^2
                 \left({R_c [{\rm pc}] }\right)^{-1}
                 \left({n_g [{\rm cm^{-3}}] }\right)^{-1} \\
               & = & 0.01246 \,\, \cdot (T_g/10^7 {\rm K})^2 
                 \cdot (R_c [{\rm pc}])^{-1} 
                 \cdot (g [\msun /{\rm pc}^3] )^{-1}  
                        \nonumber 
   \end{eqnarray}
   is smaller than 0.03, condensation occurs, otherwise the clouds
   evaporate. Note that Cowie et al. use a slightly different 
   notation: $\sigma_0$. Since the clouds have identical properties, 
   the whole system switches between condensation and evaporation,
   depending on the criterion. In a further development,
   Samland et al. (1996) consider a cloud population with a spectrum 
   of masses, and therefore one has at any time clouds that evaporate
   as well as those condensing. We discuss the implications in 
   Sect. \ref{s:samland}.

   The mass loss rate per cloud due to evaporation is from Cowie et al.:
   \begin{eqnarray}
      \label{e:massloss}
      \vert \mcdot \vert  \, [{\rm g/s}]
                 & = & 2.75 \cdot  10^4 \cdot T_g^{2.5}
                       \cdot R_c [\rm pc]  \\ 
      \vert \mcdot \vert  \, [\msun /{\rm  Myr}]    
                 & = & 1.38 \cdot 10^6  
                       \cdot (T_g/10^7 {\rm K})^{2.5} \cdot R_c [\rm pc] 
   \end{eqnarray}
   For $\sigma > 1$ the mass loss rate decreases with increasing
   $\sigma$. In what follows, we do not consider this additional
   detail, as it alters the behaviour of the models only slightly.
   This applies also to the modifications introduced by
   McKee \& Begelman (1990). From the number density of the clouds
   \begin{equation}
      \label{e:cloudnumber}
      n_c = c / M_c
   \end{equation}
   one gets the rates for evaporation and condensation 
   \begin{eqnarray}
      \label{e:erate}
      E c & = & \vert \mcdot \vert \, n_c  \\
      \label{e:krate}
      K g & = & \vert \mcdot \vert \, n_c (0.03/\sigma)   
   \end{eqnarray} 
   We note that with $\sigma = 0.03$, on the border between 
   evaporation and condensation behaviour, the rates are
   equal:
   \begin{equation}
      \label{e:switchbalance}
      Ec = Kg
   \end{equation}

   The changes in the internal energy densities are
   \begin{equation}
      \label{e:energygas}
       \egdot =   h_g s - g^2 \Lambda(T_g)
                + E c b \tilde{T}_c  - K g b T_g
                + b T_g s / \tau
   \end{equation}
   with the abbreviation $b = 1.5 \, k_{\rm B}/\mu_p = 0.013 $ in our units.
   The first term is the heating of the gas by massive stars. Because
   in our treatment of the stars the rate from continuous heating during
   the life-time of the stars $h s$ has the same dependence on $s$ as
   that from an explosive stellar death e.g. $E_{\rm SN} s / \tau$,
   processes of both types are included in the heating term:
   photoionization (with all stellar ionizing photons being absorbed by
   the gas), deposition of mechanical energy by stellar winds and 
   supernova explosions. All three processes give rate coefficients of
   the same order $h \approx 200000$. To permit a direct comparison with
   the model of K\"oppen et al. (1995) we keep $ h = 21900 $ (for
   photoionization with an efficiency of $10^{-4}$ for the conversion
   into thermal energy). A higher value gives a smaller self-regulated
   SFR and thus a longer star-formation time-scale. The second term is
   radiative cooling with a general cooling function $\Lambda(T)$
   (cf. Theis et al. 1992). In our units, and by noting the use of mass
   densities instead of number densities:
   \begin{equation}
      \label{e:cooling}
      \Lambda(T) = 1800 \cdot \sqrt{T/10000\,{\rm K}}   
   \end{equation}
   The third and fourth terms are the energy gained by the gas through
   the addition of the evaporated cloud material -- whose temperature
   shall be $\tilde{T}_c$ --  and the energy lost by the gas which is
   condensed onto the clouds. The (usually negligibly small) fifth term
   is that part of the gain if the stellar ejecta had gas temperature.
   We include this merely for convenience of analytical considerations.
   Likewise, the energy density of the cloudy medium is changed
   \begin{equation}
      \label{e:energycloud}
       \ecdot = h_c s - c^2 \Lambda(T_c) 
                 -\Psi b T_c 
                 - E c b T_c + K g b \tilde{T_g}  
   \end{equation}
   by gains from heating by massive stars and losses by radiative 
   cooling, by the losses due to matter locked up into stars 
   (usually a minor term), and by the losses due to evaporating material 
   or by gains from incorporating the condensing gas. The latter 
   temperature is designated as $\tilde{T_g}$. For simplicity, we 
   consider $h_c = h_g$.     

         \subsection{Metallicity}

   The metal masses in the gas $z_g = g Z_g$, where $Z_g$ is the
   metallicity, and the other components evolve like:
   \begin{eqnarray}
     \zgdot & = & {s \over \tau} 
           \left(\eta Z_s + y {1-\xi\eta \over \xi}(1-Z_s)\right)
           + EcZ_c - KgZ_g        \label{e:metal.gas}   \\
     \zcdot & = & -\Psi Z_c - EcZ_c + KgZ_g  
                                  \label{e:metal.clouds}\\
     \zsdot & = & \xi\Psi Z_c - sZ_s/\tau  
                                  \label{e:metal.star} \\
     \zrdot & = & (1-\xi )\Psi Z_c  + (1-\eta) s Z_s /\tau
                                  \label{e:metal.rem}
   \end{eqnarray}
   The first term in the equation for $z_g$ describes the supply to
   the ISM of metals from the massive stars. It is composed of two
   parts: the metals that where incorporated into the stars at their
   birth and are unchanged, and the metals freshly synthesized in the
   stars. For a whole stellar generation the latter is $\alpha y$,
   with the fraction $\alpha$ of the mass locked up into remnants
   which is here $\alpha = (1-\xi)+\xi(1-\eta) = 1-\xi\eta$.
   Since the yield $y$ (e.g. K\"oppen \& Arimoto 1991) refers to the
   entire stellar mass spectrum, the massive stars alone contribute
   $y/\xi$. For a primary element such as oxygen, the yield is constant,
   except for the (negligible) factor $(1-Z_s)\approx 1$ which takes
   into account that the primordial elements are used up with metal
   enrichment. For a secondary element (nitrogen) one has
   $y \propto Z_s({\rm oxygen})$. The abundances of every metal
   can be scaled to convenient reference values. In this paper,
   we divide the abundances by the yields, which are taken to be
   solar.

       \section{Models with switching between condensation and
                evaporation}
       \subsection{Experiences from full numerical solutions}
 
   This set of equations (Eqns. \ref{e:gas} to \ref{e:metal.rem}),
   which includes the switching according
   to the criterion of Cowie et al. (1981), had been incorporated
   into the chemodynamical models to compute the evolution of a
   closed-box system (Hensler \& Burkert 1991, Theis et al. 1992),
   and coupled with 1-dimensional hydrodynamics to study the evolution
   of spheroidal systems (Theis et al. 1992).
   It was found that star-formation occurs in two modes: in a continuous
   way as the consequence of self-regulation (cf. K\"oppen et al. 
   1995), but in a certain range of densities it would fluctuate very 
   strongly. These fluctuations actually are regular, large amplitude, 
   non-linear oscillations of the cloud temperature and hence of the 
   star-formation rate (cf. Fig. 7 in Theis et al. 1992). The period 
   of the order of 100 Myrs is longer than the cooling time scale of 
   the cloud gas. This indicates that the strongly damped oscillations 
   which may occur before star-formation reaches the self-regulated 
   mode (K\"oppen et al. 1995) cannot be the origin. Further inspection 
   reveals that the oscillations show up most strongly in the cloud 
   temperature and the gas density, rather weakly in the gas temperature,
   and were almost absent in the cloud density.  

   \begin{figure}
      \epsfxsize=5cm
      \epsffile{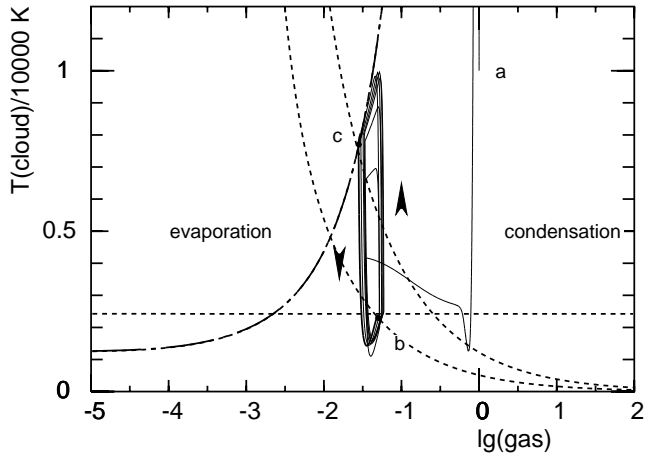}
      \caption[]{Evolution of the full system from the initial state
                 (marked `a') until the completion of the first few
                 oscillations (solid line). The dashed lines descending
                 to the right are the loci where the system switches from
                 evaporation and condensation (lower curve, at point `b')
                 and vice versa (at point `c'). The horizontal dashed line
                 is the locus of the evaporation funnel and the dot-dashed 
                 curve depicts the condensation funnel (see text).}
      \label{f:full.swing}
   \end{figure}

   In Fig. \ref{f:full.swing} we show the track in the ($g - T_c$)-plane 
   of a model entering the oscillatory mode and performing a few cycles.
   Starting from a point marked `a', it quickly and rather directly enters
   a limit cycle. At point `b' the system changes over from evaporation
   to condensation. While just a little amount of gas is condensed, the
   clouds heat up rapidly to a maximum, then cool along the dot-dashed
   curve, while much more gas condenses. This continues to point `c'
   where the system changes over to evaporation. Cooling in the clouds
   follows, with initially little gas return which increases thereafter.
   Passing through a minimum temperature, the clouds reach a temperature
   marked by the horizontal dashed line, along which the system reaches
   again point `b' and switches back to condensation. 
   As will be shown below, the horizontal dashed line is identified
   as the self-regulated star-formation mode; the passage through the
   temperature minimum is merely the transient while the system settles
   into equilibrium (K\"oppen et al. 1995).

   The two dashed curves sloping down to the right delineate the border
   where the system changes between evaporation and condensation. One
   notices that points `b' and `c' do not lie on the same curve.
   This is because the gas temperature $T_g$ at point `b' is about
   10~percent lower than at point `c'. Though this amplitude
   is rather modest, the strong dependence of the switch criterion on
   $T_g$ leads to a strong shift of the switch curve in the diagram.
   But these oscillations in $T_g$ are not essential for the behaviour
   of the model.

      \subsection{Origin of the oscillations}

   The fact that the numerical simulations show the appearance of the
   oscillations primarily in cloud temperature and gas density indicates 
   that these two variables form an `inner' system whose behaviour is 
   governed by the values of the control parameters $(c, T_g)$. 
   A most convenient way to analyze this is plotting the streamlines
   of the equations in the ($g-T_c$)-plane, with the other variables
   $c$, $s$, and $T_g$ either changing with time or held constant.
   We find that for the explanation of the fundamental structures,
   the exact dependences of the rate coefficients on density or
   temperature as given by McKee \& Begelman (1990) or Cowie et al. (1981)
   are of secondary importance. It is
   completely sufficient to assume that the rate coefficients for
   condensation and evaporation are constant ($K, E$). In the following,
   we regard this simplified description; for the criterion of
   switching between condensation and evaporation, the prescription to
   compute $\sigma$ is kept as given above, but we shall consider the
   threshold value of 0.03 as another free constant $\sigma_0$.

   \begin{figure}
      \epsfxsize=5cm
      \epsffile{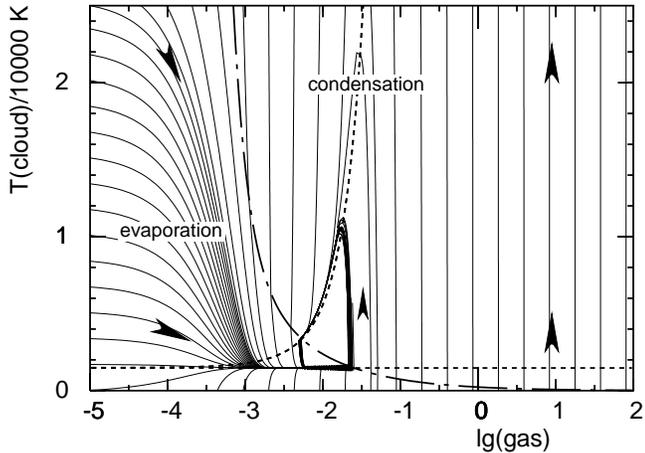}
      \caption[]{The overall behaviour of the simplified system: streamlines
         for the combined equations. The dot-dashed curve is the locus
         where the clouds change from condensation to evaporation, the short
         dashed lines mark the funnels. All streamlines below the horizontal
         dashed line point upwards.}
      \label{f:simp.all}
   \end{figure}

   The resulting streamlines are shown in Fig. \ref{f:simp.all}, which
   corresponds to Fig. \ref{f:full.swing} except for the slightly different
   position of the limit cycle. One distinguishes the two different regions
   of evaporation and condensation. In the evaporation regime, all
   streamlines are always oriented towards increasing gas density, and
   bunch together to form a funnel centered at constant
   $T_c \approx 1800$~K (corresponding to $T_c = 2400$~K in
   Fig. \ref{f:full.swing}).  Along this feature, the clouds evaporate at
   constant temperature, until the gas density exceeds a critical value
   -- corresponding to point `b' in the full model -- and condensation
   commences. Setting $K = 0$ in Eqns. \ref{e:gas}, \ref{e:cloud}, and
   \ref{e:energycloud} and using $\ecdot = b(\cldot T_c + c\tcdot)$,
   one gets the equation for the streamlines:
   \begin{eqnarray}
      {dT_c \over d\ln g} 
           & = &  g{\tcdot\over \gdot} 
           = g {h_c s - c^2 \Lambda(T_c)  
                  \over  bc(s\eta/\tau + Ec) }   \nonumber
   \end{eqnarray}
   At $g = 0$ all streamlines are horizontal, which is merely a feature
   of the logarithmic representation and gives the impression of the fan
   structure on the left as the lines bunch together into the funnel. 
   The important horizontal streamline along the funnel is described
   by the balance between stellar heating and radiative cooling 
   $h_c s = c^2 \Lambda$, i.e. the evolution proceeds with the clouds
   being essentially in thermal equilibrium. As these are the only
   features, it is evident that regardless of its initial state, the
   system will always end up in the funnel, along which it evolves at
   constant cloud temperature towards higher gas densities.  

   \begin{figure}
      \epsfxsize=5cm
      \epsffile{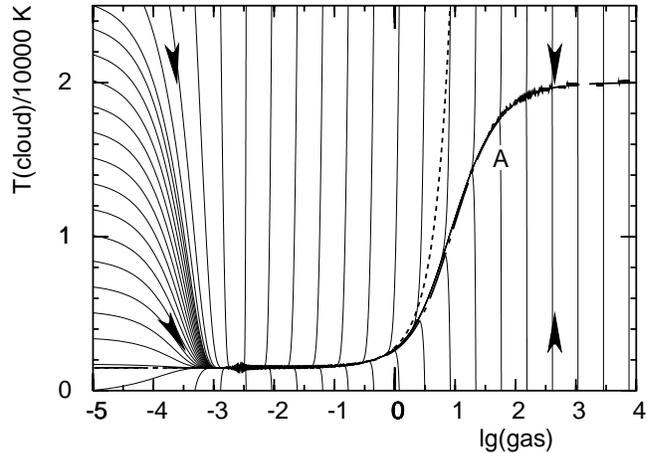}
      \caption[]{Similar to Fig. \ref{f:simp.all}, but showing the case
          for pure condensation, and assuming a temperature
          of the condensate $\tilde{T}_g = 20\,000$~K. On the dashed
          line thermal equilibrium in the clouds exists, the dot-dashed
          line shows where the cloud temperature is constant.}
      \label{f:simp.low}
   \end{figure}

   In the condensation regime, all streamlines are oriented towards
   lower gas densities. Here, one finds a funnel, marked by the
   short-dashed curve sloping down from upper right to lower left.
   All streamlines, whether coming from lower or higher temperatures,
   merge into this funnel which meets the switch line at $\lg(g) = -2.3$
   and $T_c \approx 3000$~K, and the system is changed back to
   evaporation corresponding to point `c' in Fig. \ref{f:full.swing}.
   Setting $E = 0$ yields the equation for the streamlines
   \begin{eqnarray}
      {dT_c \over d\ln g} 
          & = &  g  {(h_c s - c^2 \Lambda(T_c)) 
                      + bKg(\tilde{T}_g-T_c)
                  \over  bc(s\eta/\tau - Kg)}  \nonumber
   \end{eqnarray}
   Figure \ref{f:simp.low} depicts the whole condensation regime in the
   ($g-T_c$)-plane. To show the behaviour at large densities, a lower
   value of $\tilde{T}_g = 20\, 000$~K is assumed. Apart from a region
   of horizontal streamlines at $g = 0$, as for the evaporation region,
   one finds these structures:
   \begin{itemize}
      \item at low gas densities, a horizontal streamline exists
            where $h_c s = c^2\Lambda$, the same condition as for 
            the evaporation funnel. Its position agrees closely
            with the upper short dashed curve in Fig. \ref{f:simp.low}, 
            which is the locus where one has thermal balance in the
            clouds $\ecdot=0$.
      \item at large gas densities, a horizontal streamline appears
            where $T_c = \tilde{T}_g$. The dot-dashed curve
            (labeled `A') shows where the cloud temperature does not
            change with time. The clouds are not in thermal equilibrium,
            because the condensation occurs faster than the cooling time, 
            and the clouds are kept at the condensates' temperature. 
      \item both two parts are connected with a curve, along which 
            the evolution proceeds towards thermal equilibrium in
            the clouds. Where the transition occurs, is essentially
            determined by the ratio of $\Lambda$ and 
            $b(gK/c^2) (\tilde{T}_g - T_c)$.
      \item there is an additional feature, showing up as a clump of 
            streamlines: the intersection of the condensation funnel
            with the condition $s\eta/\tau - Kg = \gdot = 0$ at
            $\lg g = -2.6$ and $T_c = 1800$~K (in Fig. \ref{f:simp.low})
            forms an attracting  node. The evolution of the clouds
            proceeds in thermal equilibrium and in balance between
            stellar gas return and condensation. 
   \end{itemize}
   Independent of the initial conditions, the system always reaches the
   condensation funnel, where the clouds cool and the gas density
   decreases. However, under certain conditions, the system may  
   enter an equilibrium evolution at constant gas density.

   The overall evolution can thus be pieced together: The two regimes of
   evaporation and condensation are separated by the curve where the
   criterion of Cowie et al. (1981) is met exactly
   ($\sigma = \sigma_0 = 0.03$), depicted in Fig. \ref{f:simp.all}.
   One may work out how the position of this switching line depends
   on the parameters, using the prescriptions (Eqns. \ref{e:filling} 
   to \ref{e:sigma}). For a cloud filling factor much smaller than
   unity, which is the common case, one obtains:
   \begin{equation}
       \label{e:switch}
       \sigma = 4.18 \cdot 10^{-17} \cdot T_g^{2.5} \, T_c^{-1} \, g^{-0.5}
   \end{equation}
   The curve's location is independent of the cloud density $c$, and
   it is shifted towards lower gas densities, if one raises the threshold
   $\sigma_0$ or lowers the gas temperature $T_g$, which is the most
   sensitive parameter.

\begin{figure}
      \epsfxsize=5cm
      \epsffile{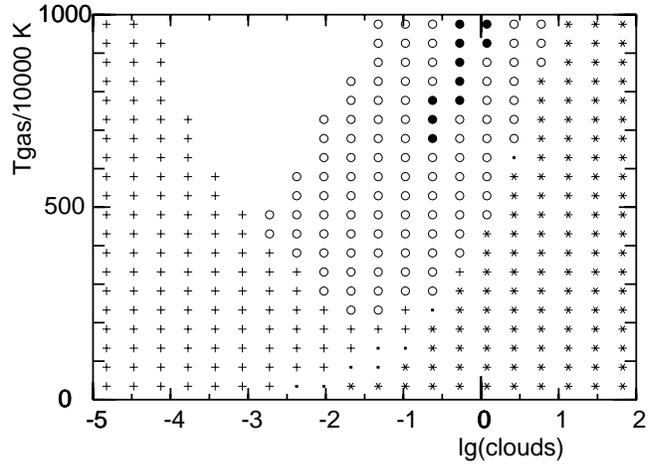}
      \caption[]{The behaviour of the simplified system as a function of the 
         control parameters $c$ and $T_g$, as estimated with the assumption
         of constant stellar density: Circles denote where oscillations
         occur (filled circles: the gas temperature is within a factor of
         2 of the value for thermal equilibrium of the gas). In the white
         area, oscillations also occur, but with a period of over 5 Gyrs. 
         Plus-signs indicate where the streamlines in both regimes are too
         flat, and the system comes to rest at an intermediate point on the
         switching line. Small dots show when the funnels meet very closely,
         and the asterisks indicate that the system enters the
         {$\gdot = 0$} equilibrium.}
      \label{f:simp.scan}
   \end{figure}

   The existence of the funnels makes it relatively easy to predict
   the conditions under which oscillations occur. From the pattern of the
   streamlines it is clear that a condition for a closed track in the
   plane is the intersection of evaporation funnel and switching
   line should happen at a lower cloud temperature (and higher gas
   density) than that of condensation funnel and switch curve. The
   conditions for the funnels and the switch line constitute a set
   of simultaneous non-linear equations which can be solved numerically
   to yield the curve in the $(g - T_c)$-plane, and the positions of
   the intersections. The stellar density can be estimated from the low
   density limit $s \approx \xi\tau\Psi$ (cf. K\"oppen et al. 1995).
   The geometry of the funnels thus allows a classification of the
   behaviour, as a function of the `outer' parameters $c$ and $T_g$,
   as shown in Fig. \ref{f:simp.scan}. Furthermore, the
   difference and ratio of the densities $g$ can be used together with
   the two forms of Eqn. \ref{e:gas} to estimate the times spent in
   each phase, the oscillation period and the keying ratio. Because these
   considerations neglect the changing gas temperature, they predict
   the types of behaviour of the complete network only in a qualitative
   way.

   All these investigations -- together with numerical solutions of
   the complete system -- show that oscillations are restricted to
   a finite range in cloud density, which may be quite narrow (less
   than a decade) and which is rather sensitive to some of the 
   parameters and constants in the model. 

   During a cycle, the following sequence of physical processes takes place:
   When condensation starts, the clouds are rapidly heated up by the gas
   condensates, until the condensation funnel is reached. There, the clouds
   cool radiatively and evolve towards thermal equilibrium. When the switching
   criterion is met again, evaporation commences. Lacking the heating by the
   condensates, the clouds cool until the new thermal equilibrium is reached.
   Then the on-going evaporation of the clouds increases the gas density,
   until the condensation criterion is again fulfilled.

   The cycle is thus essentially driven by the deposition of thermal energy
   of the condensates into the clouds, which radiate the energy away in the
   form of line emission. Thus the source which makes the cycle possible is
   the deposition of sufficiently hot material onto the clouds.
   The temperature $\tilde{T}_g$ of the condensates is an extremely
   sensitive parameter: oscillations can only be found if the condensing
   gas remains hot $\tilde{T}_g = T_g$. Assumption of a lower temperature
   greatly reduces the region for the occurrence of oscillations, or can
   completely suppress them. This is quite easily understood in terms of
   the condensation funnel (Fig. \ref{f:simp.low}): the position of its
   high-density branch is determined by the condensate temperature. Any
   cooling reduces the range in cloud temperatures that the funnel covers
   and thus the possibility for having a cyclic path between the two
   regimes is reduced. Thus, the prescription of the transfer of
   energy during the condensation phase is a most critical part of the
   network.

          \section{The physically consistent formulation}
                    \label{s:consist}

   McKee \& Cowie (1977) and McKee \& Begelman (1990) give expressions
   for the mass transfer rate, but not for the rate of energy exchange
   between gas and clouds. Therefore one considered in the previous
   chemodynamical models (e.g. Theis et al. 1992) limiting cases, such
   as that the condensing gas does not cool or cools down to some fixed
   temperature. Forming only one of the many processes of the ISM model
   network, this rather simple recipe was preferred. Of course,
   it is not a physically entirely satisfactory description, even
   more so since the oscillations were found to be rather sensitive to
   the exact recipe.

   A physically consistent prescription can be formulated from noting that 
   McKee \& Cowie (1977) as well as McKee \& Begelman (1990) derive their
   evaporation and condensation rates on the basis of a {\it stationary}
   solution for the gas flow between cloud and intercloud matter, with a
   fixed temperature profile. This means that any parcel of condensing gas
   follows the local temperature from the intercloud gas to the cloud
   interior, and that it arrives in the cloud with the cloud's temperature:
   $\tilde{T}_g = T_c$. As discussed above, cooling of the condensating
   material results in a lowering of the high-density part of the
   condensation funnel (cf. Figs. \ref{f:simp.all} and \ref{f:simp.low}).
   If now the condensates cool down to the cloud temperature, the
   condensation funnel is merely a horizontal line, at the same height
   of the evaporation funnel. Because of the opposite directions of the
   streamlines in the two regimes, the intersection of the funnels --
   which are characterized by thermal balance in the clouds -- with the
   switching curve is an attracting node, in which the system will always
   end up. Here, the system evolves with the cloud gas in thermal
   equilibrium, and without any oscillations.

   The switching condition implies that $Ec = Kg$
   (Eqn. \ref{e:switchbalance}) and so the evaporation and condensation
   terms in Eqns. \ref{e:gas}, \ref{e:cloud}, \ref{e:energygas}, and
   \ref{e:energycloud} vanish. Usually one finds that $g \ll c$, so that
   one has $c \approx c + g$, and the overall evolution is described by    
   \begin{eqnarray}
     {d \over dt} (c+g) & = &  -\Psi  + s \eta /\tau         \nonumber \\  
     \sdot     & = & \xi \Psi - s/\tau             \nonumber \\ 
     \rdot     & = & (1-\xi)\Psi + (1-\eta)s/\tau  \nonumber \\ 
     \ecdot    & = & h_c s - c^2 \Lambda(T_c) 
                         -\Psi b T_c               \nonumber 
   \end{eqnarray}
   which is nothing but the single-gas-phase system which has been shown
   to evolve almost exclusively in an equilibrium between star-formation
   and its inhibition due to the heating of the ambient gas by the 
   massive stars (K\"oppen et al. 1995).

   Note that in this more general view, the conversion of gaseous matter
   into remnants is {\it independent of the rate coefficients for
   evaporation and condensation, and even on the actual prescription for
   the switching}. But how the gas is divided among clouds and intercloud
   gas, depends on these specifications.

   The switching of the system between condensation and evaporation
   gives rise to non-continuous terms in the differential equations.
   However, since in both regimes the system evolves towards the
   switching condition, this apparent mathematical difficulty makes
   the solution instead more simple: It forces the system to
   {\it evolve along the switch line}.

   This permits to estimate the timescales of the equations: For
   conditions appropriate to a galactic disk ($c = 0.1$ \Msun /pc$^3$)
   one has $T_c \approx 3000$~K from the figures shown.
   The gas density is typically $g \approx 0.1 ... 0.01 c$.
   The cloud filling factor is usually quite small
   $\epsilon_c = cT_c/gT_g \ll 1$. From Eqn. \ref{e:switch}, one gets
   for the temperature of the intercloud gas 
   \begin{equation}
      \label{e:tgas.form}
      T_g = 8.76 \cdot 10^5 \, {\rm K} \, \cdot (\sigma_0/0.03)^{0.4}
                    \, \cdot T_c^{0.4} \, g^{0.2}
          \approx 10^7 \, {\rm K}
   \end{equation}
   which is quite insensitive to the actual gas density.
   The rate coefficients are obtained from Eqns.
   \ref{e:filling} to \ref{e:krate}
   \begin{eqnarray}
      E & = & 1.17 \cdot 10^{-17}  \cdot T_g^{2.5} \, T_c^{-1}
        \approx 0.001 \, {\rm Myr}^{-1}    \nonumber  \\ 
      K & = & E \cdot {c \over g} \approx 0.01 ... 0.1 \, {\rm Myr}^{-1}.
                           \nonumber
   \end{eqnarray} 
   This makes these processes much slower than the cooling in the clouds
   $\tau_{\rm cool,c} = b T_c / c \Lambda \approx 0.4$~Myr, but
   faster than star-formation $\tau_{\rm SFR} = c /\Psi \approx 300$~Myr.
   For larger heating coefficients than we used here, the star-formation
   is even slower (few Gyr), and the two equilibria influence each
   other even less.

   Thus, after some transient phase, the evolution proceeds in a 
   {\it hierarchy of equilibria}: the fast cooling in the clouds will 
   establish the self-regulated mode of star-formation; on a longer
   timescale the condensation and evaporation will balance the
   distribution of the gaseous matter among cloud and intercloud gas;
   and even more slowly, all gas is consumed to be turned into stellar
   remnants.

        \subsection{Balance of condensation and evaporation}

\begin{figure}
      \epsfxsize=5cm
      \epsffile{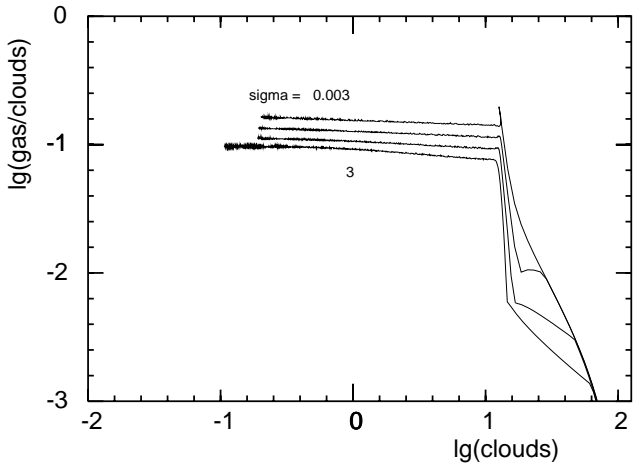}
      \caption[]{The mass ratio of gas and clouds, as a function of cloud 
          density, for different values of the switching threshold
          $\sigma_0$.}
      \label{f:gc.sigma}
   \end{figure}

   How the gas is distributed among the gas and the cloud component, 
   depends on the condensation and evaporation. As stated above,
   the system evolves along the switching line, which can be
   formulated as:
   \begin{eqnarray}
      0 = {d\ln(\sigma) \over dt} 
            =  2.5 \, {\tgdot \over T_g} - 0.5 \, {\gdot \over g}
                 - {\tcdot \over T_c}      \nonumber
   \end{eqnarray}
   Thermal equilibrium in the clouds implies constant temperature,
   so the last term vanishes, and with Eqn. \ref{e:energygas} one gets
   \begin{eqnarray}
      {\gdot \over g} = 5 \, {\tgdot \over T_g}  
                      = 5 \, {h_g s - g^2 \Lambda(T_g) \over  bgT_g } 
                                          \nonumber
   \end{eqnarray}
   The changes of the densities can be written in the low density limit 
   of self-regulated star-formation
   $s \approx \xi \tau \Psi \approx 0.03\, c^2$
   \begin{eqnarray}
      \gdot & = & a_1 \,\, g^{-0.2} (\sigma_0/0.03)^{-0.4}
                  [a_2 \,\, c^2 - a_3 \,\, g^{2.1} (\sigma_0/0.03)^{0.2}] 
                                 \nonumber   \\
     \cldot & = & -(1-\xi\eta) \Psi - \gdot \nonumber
   \end{eqnarray}
   where $T_g$ is taken from Eqn. \ref{e:tgas.form}, and using 
   the abbreviations
   \begin{eqnarray}
      a_1 & = & 1.79 \, 10^{-5} \cdot (T_c/3000 \,{\rm K})^{-0.4} \nonumber\\
      a_2 & = & 0.03            \cdot h_g \approx 660             \nonumber\\
      a_3 & = & 8.36 \, 10^4 \cdot (T_c/3000 \,{\rm K})^{0.2}     \nonumber
   \end{eqnarray}
   One notes that the equations have a critical point which obeys the 
   condition $a_2 c^2 = a_3 g^{2.1} (\sigma_0/0.03)^{0.2}$, or
   \begin{eqnarray}
      {g \over c} \propto c^{-1/21} \,\, \sigma_0^{-2/21}  \nonumber 
   \end{eqnarray}
   Thus the ratio $g/c$ due to the balance of condensation and evaporation 
   increases only very slightly as the cloud gas is consumed. It is 
   independent of the actual rate coefficients, and only weakly dependent 
   on the switching threshold. These properties are found in the numerical
   solutions, shown in Fig. \ref{f:gc.sigma}. After the transient evolution
   subsides -- within 1~Gyr, which depends on the rate coefficients -- 
   the system enters the equilibrium, where the above derived dependences
   are obeyed. This shows that, apart from transient phases, the results 
   of the chemo-dynamical models are quite insensitive to the details of
   the recipe of Cowie et al. (1981), for example the correction factor
   $\Phi$ for the influence of magnetic fields.   

   The stability of the system against perturbations of the gas
   and cloud densities and temperatures are investigated by numerical
   experiments, adding or removing a substantial (50 percent) portion
   of the gas present in the system or changing strongly the temperatures.
   After a transitory phase lasting some 100 Myrs, a new equilibrium
   solution is reached which has the same character as before, i.e.
   without oscillations. In the transient there may be a few bumps and
   wiggles, but the system never breaks into oscillations. This is as
   expected, since the disturbance moves the system in e.g. the
   $(g - T_c)$-plane to another position, from where it always evolves
   towards one of the funnels to end up in the intersection with the
   switch line. Since the character of the streamlines cannot be
   altered into the pattern responsible for oscillations, the system
   is also very robust against inflows, outflows, and external heating.
       
\subsection{Metallicity}

\begin{figure}            
      \epsfxsize=5cm          
      \epsffile{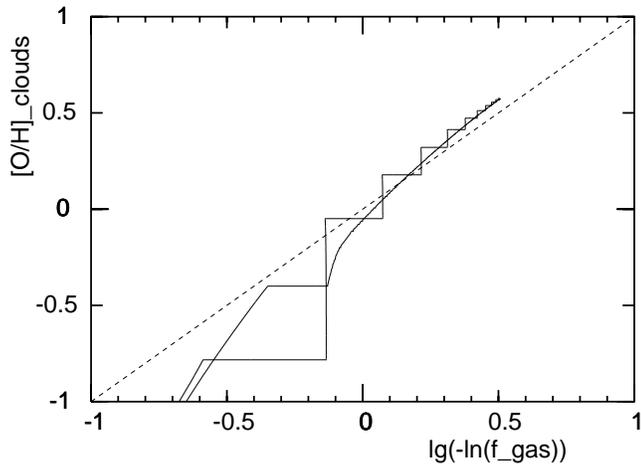}   
      \caption[]{The abundance of a primary element (oxygen) in the
               clouds as a function of the gas fraction 
               $f_{\rm gas} = (c+g)/(c+g+s+r)$. Shown are models
               with constant evaporation/condensation rate coefficients,
               assuming that the condensates retain the gas temperature
               (stepped curve) or cool down to the cloud temperature.
               The behaviour of the Simple Model is shown as a dashed line.}
     \label{f:simp.primary}
   \end{figure}

   In Fig. \ref{f:simp.primary} we depict the relation of 
   abundance in the cloud component of a primary element with the gas 
   fraction $f_{\rm gas} = (c+g) / (c+g+s+r)$. 
   In the Simple Model (closed-box with instantaneous 
   recycling) the gas metallicity follows $Z = -y\ln{f_{\rm gas}}$ 
   with the true yield $y$, shown as the dashed line.
   For simplicity, we divide the abundances by the yield.
   The abundance rises more steeply at early times than the Simple
   Model, because we allow for a finite life-time of the massive
   stars which are responsible for the metal enrichment.
       
   For comparison, we also show the model where the condensing gas does
   not cool: the metallicity remains constant during the evaporative
   phases, separated by a steep rise when the metal-rich gas is mixed to
   the clouds at the beginning of the condensation phase. If one allows
   the gas to cool down to cloud temperature before condensing, after a
   single evaporative intervall, the evolution proceeds in what seems to
   be the average evolution of the model with oscillations. It is worth
   emphasizing that for $\lg(-\ln(f_{\rm gas})) > -0.2$, the evolution 
   of either model is quite close (within 0.1 dex) to that of the  
   Simple Model, i.e. the effective yield is nearly equal to the true 
   yield.

\begin{figure}            
      \epsfxsize=5cm          
      \epsffile{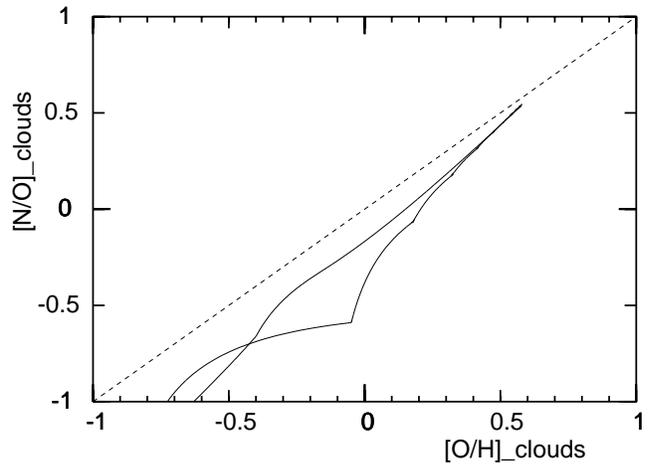}   
      \caption[]{The evolution of the abundances of a secondary (nitrogen)
              and a primary element (oxygen), for the same models shown
              in Fig. \ref{f:simp.primary}.}
     \label{f:simp.secondary}
   \end{figure}

   In Fig. \ref{f:simp.secondary} we show the behaviour of the abundance
   of a secondary element such as nitrogen, by the N/O vs. O/H plot.
   The Simple Model predicts a strictly linear relationship, viz. a
   straight line with slope 1 for the logarithmic values. Apart from
   a systematically lower yield for nitrogen (by 0.1 dex), the numerical
   models show a quite similar slope. The oscillations are less noticeable
   here than in the abundances themselves, and in the overall evolution,
   they make the slope of the N/O-O/H relation steeper. If one allows
   for cooling of the condensing gas, the slope is very close to unity.

   Thus, in both aspects, the chemical enrichment of the clouds from
   which the stars are born occurs in the very much same way as in the
   Simple Model. This should not be too surprising, because we do consider
   a closed-box model. Moreover, the time-scales for mixing of the
   freshly-produced metals into the cloud gas are determined by the
   condensation/evaporation times, which are still short compared to the
   overall evolution, viz. the gas consumption.
                                            
      \section{Models with simultaneous evaporation and condensation}
          \label{s:samland}

   The models that treat the clouds as being identical are certainly
   a simplification. For a more realistic approach, one would consider
   a cloud population with a distribution in e.g. mass, as is done
   in the models of Samland et al. (1997). With such a description
   of the cloud phase, the Cowie et al. criterion implies that
   all clouds smaller than a certain radius will evaporate, while
   those larger will grow by condensation. Thus, one has at the same
   time both processes taking place, and for the overall rates one
   integrates over the cloud spectrum. How this is done and which rates
   one gets, depends on any further assumptions on the evolution of the 
   cloud population. Samland et al. obtain for the solar neighbourhood, 
   coefficients $E \approx 10^{-3}$ Myr$^{-1}$ and
   $K \approx 0.1$ Myr$^{-1}$ which remain quite constant throughout
   the evolution.

   Such a system settles quickly, within a few condensation time-scales
   $\tau_{\rm cond}$ into the balance between condensation and
   evaporation 
   \begin{eqnarray}    \label{e:steady}
     {g \over c} = {E \over K} \nonumber
   \end{eqnarray}
   which is time-independent. The overall gas consumption takes place much
   more slowly, with $\tau_{\rm SFR}$. For a linear  SFR, this can easily
   be shown by solving the equations analytically. Since one normally has
   $\tau_{\rm SFR} > \tau_{\rm evap} > \tau_{\rm cond}$,
   most of the gaseous matter is in the cloud component, which is slowly
   converted into stellar remnants.

\begin{figure}
      \epsfxsize=5cm
      \epsffile{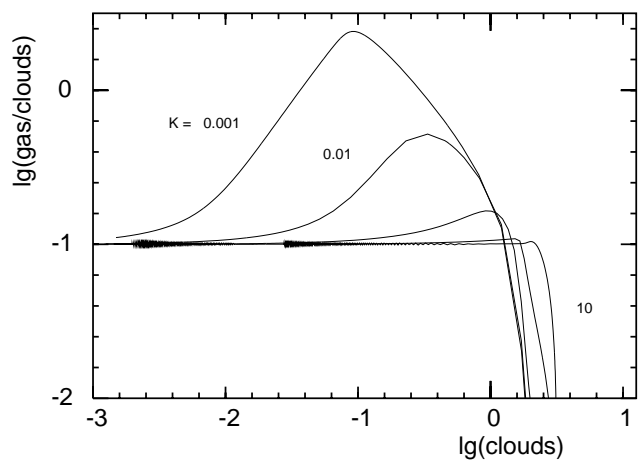}
      \caption[]{The mass ratio of gas and clouds, as a function of cloud 
          density, for different values of the condensation rate 
          coefficient $K$ (in Myr$^{-1}$), and with the same ratio
          $E/K = 0.1$.}
      \label{f:gc.trans}
   \end{figure}

   In Fig. \ref{f:gc.trans} we show the ratio $g/c$ as a function of
   cloud density for models with $E/K = 0.1$, starting with an initial
   density $c = 10^{0.5} \,\,\msun /{\rm pc}^3$. The self-regulated 
   star-formation is quickly established as soon as the star-formation
   time-scale has become larger than the mean life-time of the massive
   stars ($c/\Psi > 2\tau$ or $c < 10^{0.2}$). As long as the condensation 
   occurs faster than the stellar gas return ($K > 0.1$ Myr$^{-1}$),
   the system settles into the condensation/evaporation equilibrium.
   In models with very low condensation coefficients, the ejecta from
   the stars formed in the initial period of rather high star-formation
   are first accumulated in the hot gas, whose mass may remain constant
   while the clouds continue to be consumed -- $g/c \propto 1/c$.
   After a few condensation time-scales the gas condenses following
   the steady-state condition (Eqn. \ref{e:steady}). The peak in the 
   curve for the rather extreme model with $K = 0.001$ at $\lg c = -1$
   marks where the gas and cloud phases are in equilibrium. In this
   model, about 10 percent of the initial mass is 
   present in the gas component for about a few Gyr.  
   
\begin{figure}
      \epsfxsize=5cm
      \epsffile{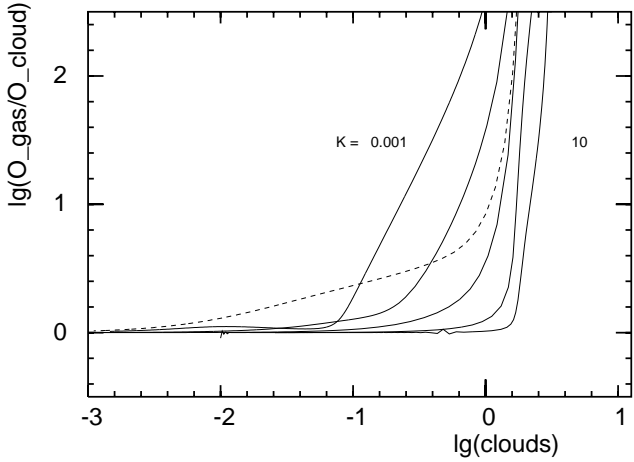}
      \caption[]{Like Fig. \ref{f:gc.trans}, but for the ratio of the 
          oxygen metallicities. The 
          dashed line is a model with $K=1$ and $E=0.01$.}
      \label{f:oxy.trans}
   \end{figure}

   In Fig. \ref{f:oxy.trans} we show the ratio of the metallicities in 
   gas and cloud components, to study the efficiency of the mixing between
   the two components. Comparing with the evolution of the density
   ratio, one notices that the metallicities become equal as soon
   as the model reaches the steady-state solution (Eqn. \ref{e:steady}).

\begin{figure}
      \epsfxsize=5cm
      \epsffile{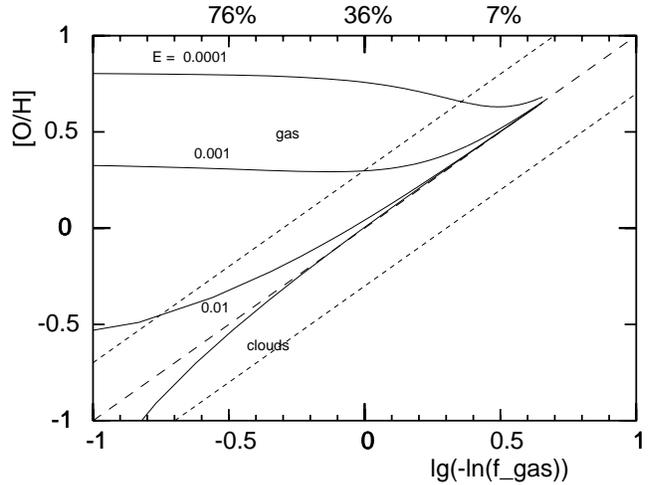}
      \caption[]{The oxygen metallicity as a function of the gas mass
          fraction, for the intercloud gas (and for different values of the
          evaporation rate coefficient $E$) and the clouds (independent of 
          $E$). The condensation coefficient is $K = 0.1$, the initial 
          cloud density is $c_0 = 0.1$ \Msun /pc$^3$. The behaviour of the 
          Simple Model with a unity yield is shown as a dashed line, the
          short dashed lines for half and double yield.}
      \label{f:yield}
   \end{figure}

   The oxygen abundance in gas and clouds as a function of the
   gas fraction is depicted in Fig. \ref{f:yield}, compared to
   the expectation from the Simple Model. If condensation 
   and evaporation occur faster than star-formation, gas and clouds   
   remain well-mixed, and the model follows the Simple Model
   very closely. The metallicity of the clouds is essentially
   determined by the amount of mixing due to condensation, it
   does not depend on the value of $E$. On the other hand, the
   gas metallicity is determined by how much the stellar ejecta
   are mixed with the low metal material evaporated from the clouds.
   For low values of $E$ the mixing is poor, and the metallicity 
   is as large as the metal abundance of the fresh stellar ejecta 
   which is as large as $y/\xi \approx 10 y$ (Eqn. \ref{e:metal.gas}). 
   Eventually, all models evolve with the true yield.
   
\begin{figure}
      \epsfxsize=5cm
      \epsffile{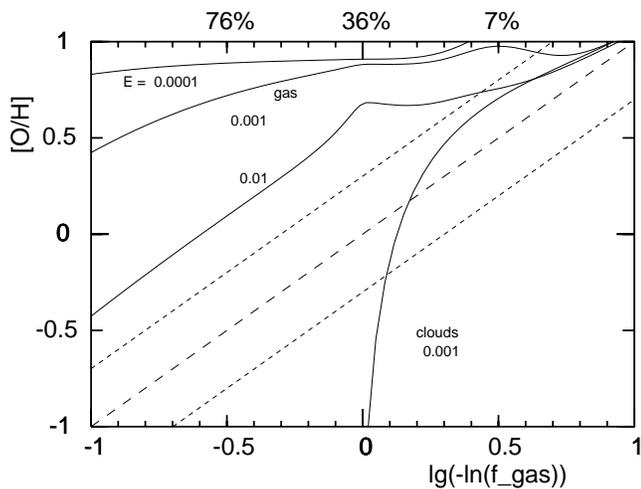}
      \caption[]{Same as Fig. \ref{f:yield}, but with a higher initial
          cloud density $c_0 = 10$ \Msun /pc$^{3}$.}
      \label{f:yield.high}
   \end{figure}

   In systems with a large initial density $c_0 \ge 1$~\Msun /pc$^3$
   (shown in Fig. \ref{f:yield.high}) the initial SFR 
   $\Psi \approx 0.03 c^2_0$ is greater than the condensation rate 
   $gK \approx (0.1 ... 0.01 \cdot c_0) \cdot 0.1$. As a consequence,
   metals accumulate in the gas, and because of the small but
   existing reprocessing of enriched cloud matter the metallicity
   even rises. The condensation of the metal-rich gas occurs only
   at a more evolved stage, and then the large gas metallicity 
   causes the clouds to have metallicities larger than expected
   from the Simple Model.                

        \subsection{Distribution of metal mass among the components}     

   In a galaxy, the total mass of a primary element such as oxygen 
   is proportional to the total number of type II supernovae that have 
   exploded until the present time, independent of when and where this 
   happened. This makes the total metal mass an essential indicator 
   for the global state of a galaxy, though it is not easy to obtain, 
   as it requires the determination of not only the mean abundances
   of gas in the various forms and stars, but their mass fractions as 
   well. This would involve different wavelength regions and abundance
   analysis techniques with the problems associated combining the data.
   
   Is it necessary to measure all gaseous components, or would the 
   observation of a single phase, e.g. the cloud components being
   observable with
   H~II regions, suffice? What is the magnitude of the errors involved?
   In the following, we use our closed-box chemodynamical models to
   provide some rough answers. Since the freshly produced metals are first 
   injected into the warm or hot intercloud gas, observations of
   the dense cloud medium might miss an important fraction of the metals.

\begin{figure}
      \epsfxsize=5cm
      \epsffile{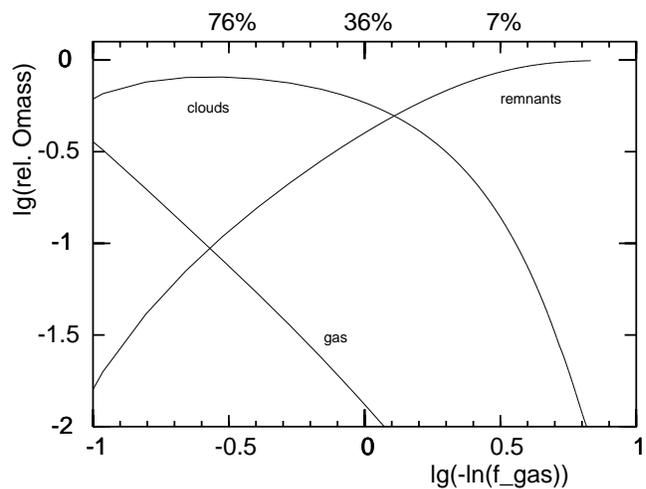}
      \caption[]{The fraction of the oxygen mass in the gas, clouds,
          and remnants, as a function of the gas mass fraction $f_{\rm gas}$.
          Model parameters are $c_0 = 0.1$ \Msun /pc$^3$, $K = 0.1$~Myr$^{-1}$,
          and $E/K = 0.01$.}
      \label{f:omassa.frac}
   \end{figure}

\begin{figure}
      \epsfxsize=5cm
      \epsffile{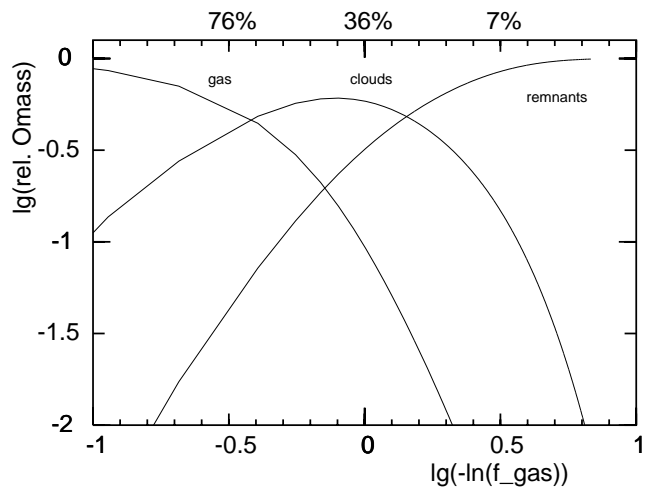}
      \caption[]{Same as Fig.\ref{f:omassa.frac} but with
         $K = 0.01$~Myr$^{-1}$.}
      \label{f:omassb.frac}
   \end{figure}

\begin{figure}
      \epsfxsize=5cm
      \epsffile{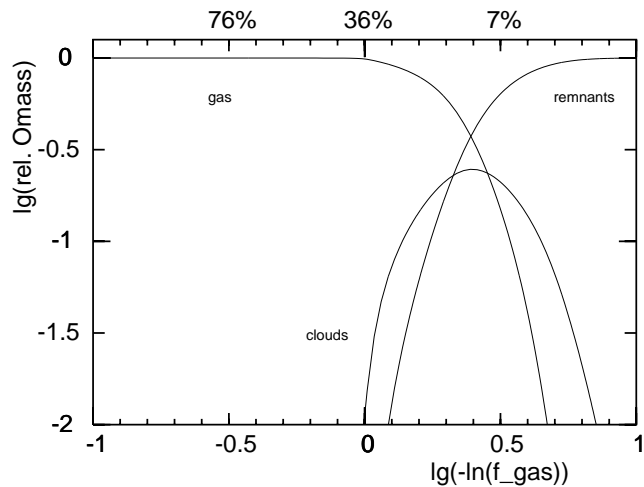}
      \caption[]{Same as Fig.\ref{f:omassb.frac} but with
         an initial density $c_0 = 10$ \Msun /pc$^3$.} 
      \label{f:omassc.frac}
   \end{figure}

   In Figs. \ref{f:omassa.frac} to \ref{f:omassc.frac} we show 
   how the distribution of oxygen mass over gas, clouds, and remnants
   depends on the gas mass fraction $f_{\rm gas}$, which indicates the
   chemical age of a system. The metals in chemically very young systems 
   are still mainly in the hot gas. At more advanced stages, the metal
   rich gas has condensed into the clouds, which then contain most
   metal mass. In highly evolved systems, the metals are found in the
   remnants. The shape of the curves in intermediately evolved systems,
   i.e. around the peak of the contribution from the clouds, is 
   mainly determined by the ratio of the inverse time-scales for 
   condensation and initial star formation 
   $K / (\Psi/c_0) \propto K \cdot c_0$.  
   Because in the closed-box model most of the star formation occurs
   during the initial phase, this time-scale determines how much metals
   are produced and which are released from the intercloud medium by  
   condensation $K$. For  $K \cdot c_0 < 1$ \Msun pc$^{-3}$ Myr$^{-1}$ 
   (Fig. \ref{f:omassa.frac}) the metals are hardly to be found in the gas, 
   because of the rapid condensation into clouds. For values smaller than 
   about 0.1 (Fig. \ref{f:omassc.frac})
   the metals remain in the gas until rather late, and because of the
   fast conversion of clouds into stars, they are swiftly locked up into
   remnants. As the clouds may contain less than a third of all the metals,
   measurements of the metallicity of the cloud gas only
   could be rather misleading for the determination of the total
   metal mass in a galaxy. Such a situation arises for low condensation
   rates or for high initial gas densities, i.e. a centrally rather 
   concentrated proto-galactic cloud. 

   The shapes of the curves are much less strongly influenced by the
   ratio $E/K$ which we kept constant in the figures: It determines the 
   ratio of metal mass contained in gas and clouds in the late evolution; 
   thus because one has $E < K$, the clouds keep a larger portion of the 
   metal mass than the gas. 

             \section{Conclusions}

   We investigate the behaviour of one-zone chemical evolution models 
   which utilize the chemodynamical prescription for a two-component 
   interstellar medium. This is done by numerical solution, topological 
   analysis of the equations, and identification of the conditions for 
   equilibrium solutions.
   
   The self-regulated star-formation found in models with a single
   gas component (K\"oppen et al. 1995) is not at all disturbed
   by the additional condensation and evaporation processes. Due
   to its shorter time-scale, it dominates the evolution. The models
   follow the self-regulated solution very closely, without any oscillations.               
   
   On the other hand, it is the distribution of the gas among the cloud 
   and intercloud phases which is determined by the mass exchange due
   to condensation and evaporation. The evolution also follows an 
   equilibrium which is either ruled by the switching condition 
   $\sigma = \sigma_0$ or by the relation $gK = Ec$. The mass of gas in 
   the intercloud phase is usually very small compared to that found in 
   clouds.
     
   From realistic rate coefficients for condensation and evaporation
   in a cloud population (Samland et al. 1997) one finds that the 
   evolution of the model follows a hierarchy of nested equilibria, 
   which are in ascending order of their quite separate time-scales: 
   thermal equilibrium in the cloud gas, regulation of the SFR by 
   heating of gas due to radiation from massive stars, equilibrium 
   of mass exchange between gas and clouds. This is further embedded 
   into the overall conversion of gas into stellar remnants.

   During the evolution in the equilibrium between condensation
   and evaporation $gK = Ec$, the relation between metallicity and gas
   mass fraction follows a Simple Model quite closely. Gas and cloud 
   phases are well-mixed, their metallicities are the same.

   Together with the quadratic dependence of the star-formation rate
   on the cloud density (cf. K\"oppen et al. 1995), these aspects
   make up the global behaviour that is typical for chemo-dynamical
   evolution models.

   In models with low condensation coefficients and/or high initial
   cloud densities, the metal-rich gas ejected by the stars formed
   in the initial period of intense star-formation may remain in the
   intercloud medium quite long, before condensation into clouds commences
   and metal enrichment of the clouds occurs. 

   From these models we compute how the total metal mass in a galaxy
   is distributed among gas, clouds, and stellar remnants and how this
   changes with the chemical age of the system. For conditions similar
   to the solar neighbourhood, unevolved gas-rich systems have their 
   metals in both cloud and intercloud gas. 
   In high-density models (or those with small condensation rates) 
   the metals may remain in the intercloud medium even in rather evolved,
   gas-rich galaxies, and those contained in the cloud phase may be a 
   rather portion. Work is under way to investigate the properties
   of full `chemodynamical' models with the complete network of
   interactions and the coupling to the global dynamics.

{\em Acknowledgements:} We thank M. Samland and R. Spurzem for 
      valuable discussions. J.K. thanks the Deutsche 
      Forschungsgemeinschaft for financial support (project He 1487/13-1) 
      and the Observatoire de Strasbourg for its generous hospitality.

\end{document}